\documentclass{article}
\usepackage{ismir,amsmath,cite,url}
\usepackage{graphicx}
\usepackage{color}
\usepackage{makecell}
\usepackage{multirow}
\usepackage{booktabs}
\title{A data-driven approach to mid-level perceptual musical feature modeling}

\threeauthors
  {Anna Aljanaki} {Institute of Computational Perception, \\ Johannes Kepler University \\ {\tt aljanaki@gmail.com}}
  {} {\bf \\\bf }
  {Mohammad Soleymani} {Swiss Center for Affective Sciences, \\Geneva University \\ {\tt mohammad.soleymani@unige.ch}}

\begin{document}

\maketitle
\begin{abstract}

Musical features and descriptors could be coarsely divided into three levels of complexity. The bottom level contains the basic building blocks of music, e.g., chords, beats and timbre. The middle level contains concepts that emerge from combining the basic blocks: tonal and rhythmic stability, harmonic and rhythmic complexity, etc. High-level descriptors (genre, mood, expressive style) are usually modeled using the lower level ones. The features belonging to the middle level can both improve automatic recognition of high-level descriptors, and provide new music retrieval possibilities. Mid-level features are subjective and usually lack clear definitions. However, they are very important for human perception of music, and on some of them people can reach high agreement, even though defining them and therefore, designing a hand-crafted feature extractor for them can be difficult. In this paper, we derive the mid-level descriptors from data. We collect and release a dataset\footnote{https://osf.io/5aupt/} of 5000 songs annotated by musicians with seven mid-level descriptors, namely, melodiousness, tonal and rhythmic stability, modality, rhythmic complexity, dissonance and articulation. We then compare several approaches to predicting these descriptors from spectrograms using deep-learning. We also demonstrate the usefulness of these mid-level features using music emotion recognition as an application.

\end{abstract}
\section{Introduction}\label{sec:introduction}

In music information retrieval, features extracted from audio or a symbolic representation are often categorized as low or high-level \cite{Casey2008}, \cite{Pampalk2006}. There is no clear boundary between these concepts and the terms are not used consistently. Usually, features that were extracted using a small analysis window that does not contain temporal information are called low-level (e.g., spectral features, MFCCs, loudness). Features that are defined within a longer context (and often related to music theoretical concepts) are called high-level (key, tempo, melody). In this paper, we will look at these levels from the point of view of human perception, and define what constitutes low, middle and high levels depending on complexity and subjectivity of a concept. Unambiguously defined and objectively verifiable concepts (beats, onsets, instrument timbres) will be called low-level. Subjective, complex concepts that can only be defined by considering every aspect of music will be called high-level (mood, genre, similarity). Everything in between we will call mid-level. 

Musical concepts can best be viewed and defined through the lens of human perception. It is often not enough to approximate them through a simpler concept or feature. For instance, music speed (whether music is perceived as fast or slow) is not explained by or equivalent to tempo (beats per minute). In fact, perceptual speed is better approximated (but not completely explained) by onset rate \cite{Friberg2011}. There are many examples of mid-level concepts: harmonic complexity, rhythmic stability, melodiousness, tonal stability, structural regularity \cite{gabrielsson2001}, \cite{Wedin1972}. Such meta language could be used to improve search and retrieval, to add interpretability to the models of high-level concepts, and may be even break the glass ceiling in the accuracy of their recognition.

In this paper we collect a dataset and model these concepts directly from data using transfer learning.

\begin{table*}[ht]
 \centering
 \begin{tabular}{l|c|c}
\toprule
Perceptual Feature & Criteria when comparing two excerpts & Cronbach's $\alpha$\\
\toprule
Melodiousness  & To which excerpt do you feel like singing along? & 0.72\\
\midrule
Articulation & Which has more sounds with staccato articulation? &  0.8\\
\midrule
Rhythmic stability & \makecell{Imagine marching along with music. \\Which is easier to march along with? }& 0.69\\ 
\midrule
Rhythmic complexity & \makecell{Is it difficult to repeat by tapping? \\ Is it difficult to find the meter? \\ Does the rhythm have many layers?} & 0.27 (0.47) \\
\midrule
Dissonance & \makecell{ Which excerpt has noisier timbre? \\ Has more dissonant intervals (tritones, seconds, etc.)? } & 0.74 \\
\midrule
Tonal stability & \makecell{ Where is it easier to determine the tonic and key? \\ In which excerpt are there more modulations? } & 
0.44 \\
\midrule
Modality & \makecell{ Imagine accompanying this song with chords. \\ Which song would have more minor chords? } & 0.69 \\
\bottomrule
 \end{tabular}
 \caption{Perceptual mid-level features and the questions that were provided to raters to help them compare two excerpts.}
 \label{tab:midlevel_features}
\end{table*}

\section{Related work}

Many algorithms have been developed to model features describing such aspects of music as articulation, melodiousness, rhythmic and dynamic patterns. MIRToolbox and Essentia frameworks offer many algorithms that can extract features related to harmony, rhythm, articulation and timbre \cite{Lartillot2008}, \cite{Bogdanov2013}. These features are usually extracted using some hand-crafted algorithm and have a differing amount of psychoacoustic and perceptual basis.

For example, Salamon et al. developed a set of melodic features which extract pitch contours from a melody obtained with a melody extraction algorithm \cite{Salamon2012}. There were proposed measures like percussiveness {\cite{Pampalk2006}}, pulse clarity \cite{Lartillot2008b}, danceability \cite{streich2005}. Panda et al. proposed a set of algorithms to extract descriptors related to melody, rhythm and texture from MIDI and audio \cite{Panda2018}. It is out of our scope to review all existing algorithms for detecting what we call mid-level perceptual music concepts. 

All the algorithms listed so far were designed with some hypothesis about music perception in mind. For instance, Essentia offers an algorithm to compute sensory dissonance, which sums up the dissonance values for each pair of spectral peaks, based on dissonance curves obtained from perceptual measurements \cite{plomp1965}. Such an algorithm measures a specific aspect of music in a transparent way, but it is hard to say, whether it captures all the aspect of a perceptual feature.

Friberg et al. collected perceptual ratings for nine features (rhythmic complexity and clarity, dynamics, harmonic complexity, pitch, etc.) for a set of 100 songs and modeled them using available automatic feature extractors, which showed that algorithms can cope with some concepts and fail with some others \cite{Friberg2011}. For instance, for such an important feature like modality (majorness) there is no adequate solution yet. It was also shown that with just several perceptual features it is possible to model emotion in music with a higher accuracy than it is possible using features, extracted with MIR software \cite{Aljanaki2014}, \cite{Friberg2011}, \cite{Friberg2014}.

In this paper we propose an approach to mid-level feature modeling that is more similar to automatic tagging \cite{Choi2016}. We try to approximate the perceptual concepts by modeling them straight from the ratings of listeners. 

\section{Data collection}\label{sec:page_size}

From the literature (\cite{gabrielsson2001}, \cite{Wedin1972}, \cite{Friberg2011}) we composed a list of perceptual musical concepts and picked 7 recurring items. \tabref{tab:midlevel_features} shows the selected terms. The concepts that we are interested in stem from musicological vocabulary. Identifying and naming them is a complicated task that requires musical training. This doesn't mean that these concepts are meaningless and are not perceived by an average music listener, but we can not trust an average listener to apply the terms in a consistent way. We used Toloka\footnote{toloka.yandex.ru} crowd-sourcing platform to find people with musical training to do the annotation. We invited anyone who has music education to take a musical test, which contained questions on harmony (tonality, identifying mode of chords), expressive terms (rubato, dynamics, articulation), pitch and timbre. Also, we asked the crowd-sourcing workers to shortly describe their music education. From 2236 people who took the test slightly less than 7\% (155 crowd sourcing workers) passed it and were invited to participate in the annotation.

\begin{figure*}
  \includegraphics[width=\textwidth]{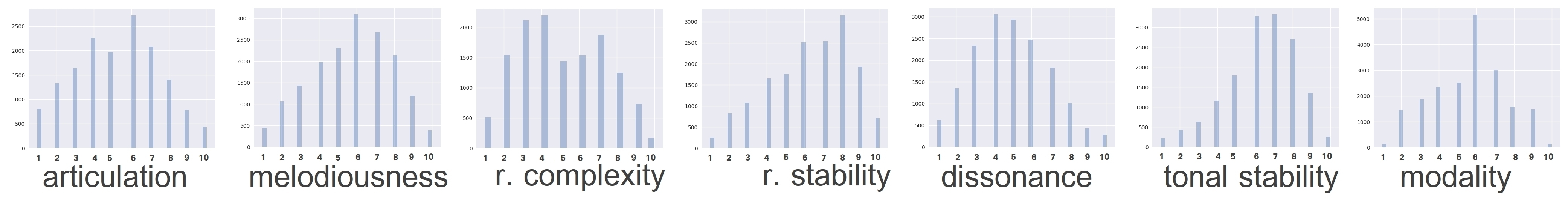}
  \caption{Distribution of discrete ratings per perceptual feature.}
  \label{fig:hist_features}
\end{figure*}

\begin{table*}[!ht]
\small
 \centering
 \begin{tabular}{lrrrrrr}
\toprule
 Feature &  Articulation & R. comlexity & R. Stability & Dissonance & Tonal stability & Mode \\
\midrule
Melodiousness	& $-0.13$ & $-0.22$ & $0.27$ & $-0.59$ &$ \mathbf{0.58}$ &$-0.22$ \\
Articulation	&         & $0.39$ & $ \mathbf{0.60}$  & $0.45$ & $-0.05$&$-0.14$ \\
R. complexity	&         &  & $-0.009$ & $0.48$ & $-0.30$ & $0.06$ \\
R. stability	& 	     &  &  & $0.06$ & $0.36$& $-0.17$ \\
Dissonance      &        &  &  &  & $ \mathbf{-0.55}$& $0.23$\\
Tonal stability	&        &  &  &  & & $-0.16$\\
\bottomrule
 \end{tabular}

 \caption{Correlations between the perceptual mid-level features.}

 \label{tab:correlations_midfeatures}
\end{table*}

\subsubsection{Definitions}

The terminology (articulation, mode, etc.) that we use is coming from musicology, but it was not designed to be used in a way that we use it. For instance, the concept of articulation is defined for a single note (or can also be extended to a group of notes). Applying it to a real-life recording with possibly several instruments and voices is not an easy task. To ensure common understanding, we offer the annotators a set of definitions as shown in \tabref{tab:midlevel_features}. The general principle is to consider the recording as a whole. 

\subsection{Pairwise comparisons}

It is easier for annotators to compare two items using a certain criterion, then to give a rating on an absolute scale, and especially so for subjective and vaguely defined concepts \cite{Madsen2013}. Then, a ranking can be formed from pairwise comparisons. However, annotating a sufficient amount of songs using pairwise comparisons is too labor intensive. Collecting a full pairwise comparison matrix (not counting repetitions and self-similarity) requires $(n^2 - n)/2$ comparisons. For our desired target of 5000 songs, that would mean $\approx12.5$ million comparisons. It is possible to construct a ranking with less than a full pairwise comparison matrix, but still for a big dataset it is not a feasible approach. We combine the two approaches. In order to do that, we first collected pairwise comparisons for a small amount of songs, obtained a ranking, and then created an absolute scale that we used to collect the rankings. 

In this way, we also implicitly define our concepts through examples without a need to explicitly describe all their aspects. 

\subsubsection{Music selection}
\label{sec:pairwise_comparisons}

For pairwise comparisons, we selected 100 songs. This music needed to be diverse, because it was going to be used as examples and needed to be able to represent the extremes. We used 2 criteria to achieve that - genre and emotion. From each of the 5 music preference clusters of Rentfrow et al. \cite{Rentfrow2011} we selected a list of genres belonging to these clusters and picked songs from the DEAM dataset \cite{Aljanaki2017} belonging to these genres (pop, rock, hip-hop, rap, jazz, classical, electronic), taking 20 songs from each of the preference clusters. Also, using the annotations from DEAM, we assured that the selected songs are uniformly distributed over the four quadrants of valence/arousal plane. From each of the songs we cut a segment of 15 seconds. 

For a set of 100 songs we collected 2950 comparisons. Next, we created a ranking by counting the percentage of comparisons won by a song relative to an overall number of comparisons per song. By sampling from that ranking we created seven scales with song examples from 1 to 9 for each of the mid-level perceptual features (for instance, from the least melodious (1) to the most melodious (9)). Some of the musical examples appeared in several scales. 

\subsection{Ratings on 7 perceptual mid-level features}

The ratings were again collected on Toloka platform, and the workers were selected using the same musical test. The rating procedure was as follows. First, a worker listened to a 15-second excerpt. Next, for a certain scale (for instance, articulation), a worker compared an excerpt with examples arranged from "legato" to "staccato" and found a proper rating. Finally, this was repeated for each of the 7 perceptual features. 

\begin{table*}[ht]
 \centering
 \begin{tabular}{l|c|c}
\toprule
\makecell{Emotional dimension \\ or category } & \makecell{Pearson's $\rho$ \\(prediction)} &  \makecell{Important features}\\
\toprule
Valence  & 0.88  & Mode (major), melodiousness (pos.), dissonance (neg.) \\
\midrule
Energy & 0.79 &  Articulation (staccato), dissonance (pos.)\\
\midrule
Tension & 0.84 & Dissonance (pos.), melodiousness (neg.)\\ 
\midrule
Anger &  0.65 & Dissonance (pos.), mode (minor), articulation (staccato) \\
\midrule
Fear  &0.82 & Rhythm stability (neg.), melodiousness (neg.)\\
\midrule
Happy & 0.81 & Mode (major), tonal stability (pos.) \\
\midrule
Sad & 0.73 & Mode (minor), melodiousness (pos.) \\
\midrule
Tender & 0.72 & Articulation (legato), mode (minor), dissonance (neg.)\\
\bottomrule
 \end{tabular}
 \caption{Modeling emotional categories in Soundtracks dataset using seven mid-level features.}
 \label{tab:performance_on_soundtracks}
\end{table*}

\subsubsection{Music selection}

Most of the dataset music consists of Creative Commons licensed music from \url{jamendo.com} and \url{magnatune.com}. For annotation, we cut 15 seconds from the middle of the song. In the dataset, we provide the segments and the links to a full song. There is a restriction of no more than 5 songs from the same artist. 
The songs from \url{jamendo.com} were also filtered by popularity, in a hope to get music of a better recording quality. We also reused the music from datasets annotated with emotion \cite{eerola2011}, \cite{panda2013}, \cite{malheiro2016} which we are going to use to indirectly test the validity of the annotations. 

\subsubsection{Data}

Figure \ref{fig:hist_features} shows the distributions of the ratings for every feature. The music in the dataset leans slightly towards being rhythmically stable, tonally stable and consonant. The scales could be also readjusted to have more examples in the regions of the most density. That might not necessarily help, because the observed distributions could also be the artifacts of people prefering to avoid the extremes. Table \ref{tab:correlations_midfeatures} shows the correlation between different perceptual features. There is a strong negative correlation between melodiousness and dissonance, a positive relationship between articulation and rhythmic stability. Tonal stability is negatively correlated with dissonance and positively with melodiousness.

\subsection{Consistency}

Any crowd-sourcing worker could stop annotating at any point, so the amount of annotated songs per person varied. An average amount of songs per worker was $187.01\pm500.68$. On average, it took $\approx2$ minutes to answer all the seven questions for one song. Our goal was to collect 5 annotations per song, which amounts to $\approx833$ man-hours. In order to ensure quality, a set of songs with high quality annotations (high agreement by well-performing workers) was interlaced with new songs, and the annotations of every crowd-sourcing worker were compared against that golden standard. The workers that gave answers very far from the standard were banned. Also, the answers were compared to the average answer per song, and workers whose standard deviation was close to one one resulting from random guessing were also banned and their answers discarded. The final annotations contain answers of 115 workers out of a pool of 155, who passed the musical test.

\tabref{tab:midlevel_features} shows a measure of agreement (Cronbach's $\alpha$) for each of the mid-level features. The annotators reach good agreement for most of the features, except rhythmic complexity and tonal stability. We created a different musical test, containing only questions about rhythm, and collected more annotations. Also, we provided more examples on the rhythm complexity scale. It helped a little (Cronbach's $\alpha$ improved from 0.27 to 0.47), but still rhythmic complexity has much worse agreement than other properties. In a study of Friberg and Hedblad \cite{Friberg2011}, where similar perceptual features were annotated for a small set of songs, the situation was similar. The least consistent properties were harmonic complexity and rhythmic complexity. 

We average the ratings for every mid-level feature per song. The annotations and the corresponding excerpts (or links to external reused datasets) are available online (osf.io/5aupt). All the experiments below are performed on averaged ratings. 

\begin{table}[ht]
 \centering
 \begin{tabular}{l|c|c}
\toprule
\makecell{Cluster } & AUC & F-measure\\
\toprule
\makecell{Cluster 1 \\passionate, confident}& 0.62 & 0.38\\
\midrule
\makecell{Cluster 2 \\cheerful, fun}& 0.7 & 0.5 \\
\midrule
\makecell{Cluster 3 \\ bittersweet}& 0.8 & 0.67 \\ 
\midrule
\makecell{Cluster 4\\ humorous}& 0.65 & 0.45 \\
\midrule
\makecell{Cluster 5 \\aggressive}& 0.78 & 0.64 \\
\bottomrule
 \end{tabular}

 \caption{Modeling MIREX clusters with perceptual features.}
 \label{tab:performance_on_multimodal}
\end{table}

\begin{figure*}
  \includegraphics[width=\textwidth]{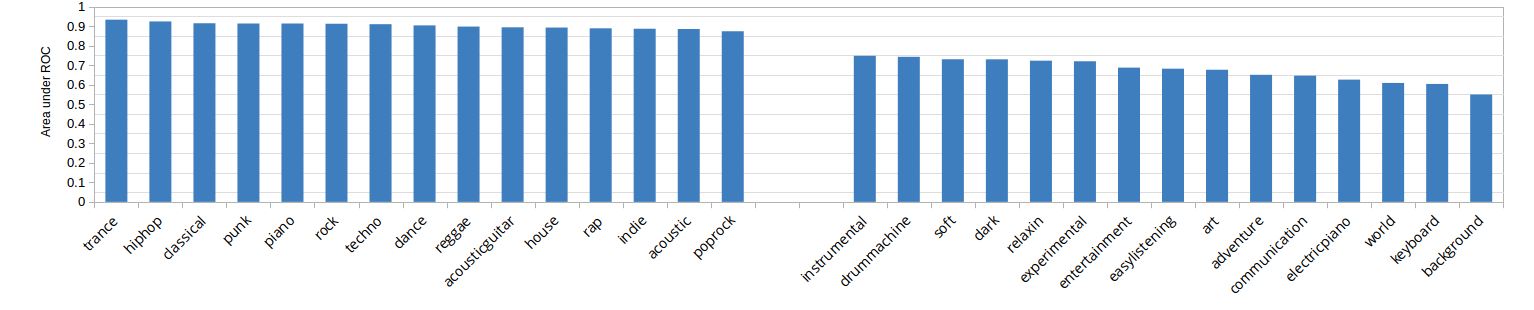}
  \caption{AUC per tag on the test set.}
  \label{fig:result_tags}
\end{figure*}

\subsection{Emotion dimensions and categories}

Soundtracks dataset contains 15 second excerpts from film music, annotated with valence, arousal, tension, and 5 basic emotions \cite{eerola2011}. 

We show that our annotations are meaningful by using them to model musical emotion in Soundtracks dataset. The averaged ratings per song for each of the seven mid-level concepts are used as features in a linear regression model (10-fold cross-validation). 

Table \ref{tab:performance_on_soundtracks} shows the correlation coefficient and the most important features for each dimension, which are consistent with the findings in the literature \cite{gabrielsson2001}. We can model most dimensions well, despite not having any information about loudness and tempo.

\subsection{MIREX clusters}

Multimodal dataset contains 903 songs annotated with 5 clusters used in MIREX Mood recognition competition \footnote{www.music-ir.org/mirex} \cite{panda2013}. Table \ref{tab:performance_on_multimodal} shows results of predicting the five clusters using the seven mid-level features and an SVM classifier. The average weighted F1 measure on all the clusters on this dataset is 0.54. In \cite{panda2013}, with an SVM classifier trained on 253 audio features, extracted with various toolboxes, F1 measure was 44.9, and 52.3 with 98 melodic features. By combining these feature sets and doing feature selection by using feature ranking, the F1 measure was increased to 64.0. Panda et al. hypothesize that Multimodal dataset is more difficult than MIREX dataset (their method performed better (0.67) in MIREX competition than on their own dataset). In MIREX data, the songs went through an additional annotation step to ensure agreement on cluster assignment, and only songs that 2 out of 3 experts agreed on were kept. 

\section{Experiments}\label{sec:experiments}

We left out 8\% of the data as a test set. We split the train set and test set by performer (no performer from the test set appears in the training set). Also, all the performers in the test set are unique. For pretraining, we used songs from \url{jamendo.com}, making sure that the songs used for pretraining do not reappear in the test set. The rest of the data was used for training and validation (whenever we needed to validate any hyperparameters, we used 2\% of the train set for that). 

From each of the 15-second excerpts we computed a mel-spectrogram with 299 mel-filters and a frequency range of 18000Hz, extracted with 2048 sample window (44100 sampling rate) and a hop of 1536. In order to use it as an input to a neural network, it was cut to a rectangular shape (299 by 299) which corresponds to about 11 seconds of music. Because the original mel-spectrogram is a bit larger, we can randomly shift the rectangular window and select a different set. For some of the songs, full-length songs are also available, and it was possible to extract the mel-spectrogram from any place in a song, but in practice this worked worse than selecting a precise spot. 

We also tried other data representations: spectrograms and custom data representations (time-varying chroma for tonal features and time-varying bark-bands for rhythmic features). Custom representations were trained with a two-layer recurrent network. These representations worked worse than mel-spectrograms with a deep network.

\subsection{Training a deep network}

We chose Inception v3 architecture \cite{Szegedy2016}. First five layers are convolutional layers with 3 by 3 filters. Twice max-pooling is applied. The last layers of the network are the so-called "inception layers", which apply filters of different size in parallel and merge the feature maps later. We begin by training this network without any pretraining.

\subsubsection{Transfer learning}

With a dataset of only 5000 excerpts, it is hard to prevent overfitting when learning features from the very basic music representation (mel-spectrogram), as it was done in \cite{Choi2016} on a much larger dataset. In this case, transfer learning can help.

\subsubsection{Data for pretraining}

We crawl data and tags from Jamendo, using the API provided by this music platform. We select all the tags, which were applied to at least 3000 songs. That leaves us with 65 tags and 184002  songs. For training, we extract a mel-spectrogram from a random place in a song. We leave 5\% of the data as a test set. After training on mini-batches of 32 examples with Adam optimizer for 29 epochs, we achieve an average area under receiver-operator curve of 0.8 on the test set. The AUC on the test set grouped by tag are shown on Figure \ref{fig:result_tags} (only 15 best and 15 worst performing tags). Some of the songs in the mid-level feature dataset also were chosen from Jamendo.

\begin{figure*}
  \includegraphics[width=\textwidth,height=5cm]{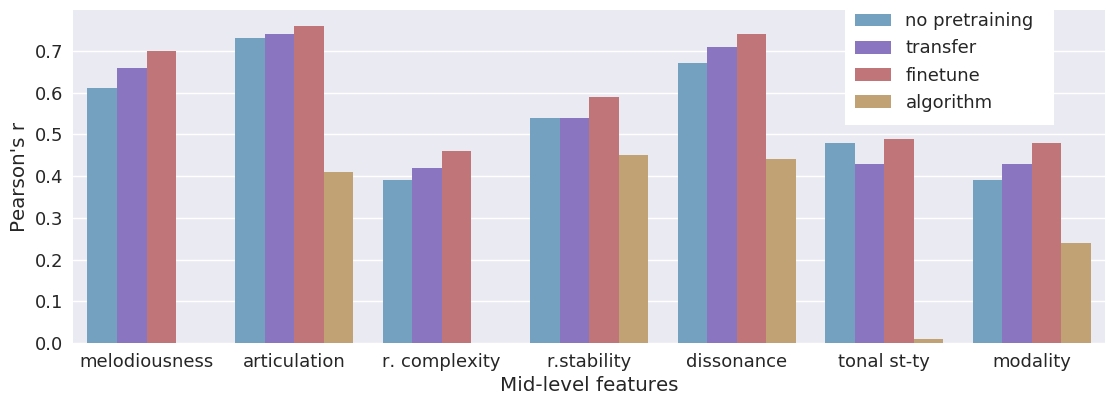}
  \caption{Performance of different methods on mid-level feature prediction.}
  \label{fig:results_features}
\end{figure*}

\subsubsection{Transfer learning on mid-level features}

The last layer of Inception, before the 65 neurons that predict classes (tags), contains 2048 neurons. We pass through the mel-spectrograms of the mid-level feature dataset and extract the activations of this layer. We normalize these extracted features using mean and standard deviation of the training set. On the training set, we fit a PCA with 30 principle components (the number was chosen based on decline of eigenvalues of the components) and then apply the learned transformation on a validation and test set. On a validation set, we tune parameters of a SVR with a radial basis function kernel and finally, we predict the seven mid-level features on the test set. 

\subsection{Fine-tuning trained model for mid-level features}

On top of the last Inception layer we add two fully-connected layers with 150 and 30 neurons, both with ReLU activation, and an output layer with 7 nodes with no activation (we train on all the features at the same time). First, we freeze the pre-trained weights of the Inception and train the last layer weights until there's no improvement on the validation set anymore. At this point, the network reaches the same performance on the test set as it reached using transfer learning and PCA (which is what we would expect). Now, we unfreeze the weights and with a small learning rate continue training the whole network until it stops improving on validation set. 

\subsection{Existing algorithms}

There are many feature extraction frameworks for MIR. Some of those (jAudio, Aubio, Marsyas) only offer timbral and spectral features, others (Essentia, MIRToolbox, VAMP Plugins for Sonic Annotator) offer features, which are similar to the mid-level features of this paper. Figure \ref{fig:results_features} shows the correlation of some of these features with our perceptual ratings: 

\begin{enumerate}
\item  \emph{Articulation}. MIRToolbox offers features describing characteristics of onsets (attack time, attack slope, leap (duration of attack), decay time, slope and leap. Out of this features leap was chosen (it had the strongest correlation with perceptual articulation feature). 
\item \emph{Rhythmic stability}. Pulse clarity (MIRToolbox) \cite{olartillot2008}.
\item \emph{Dissonance}. Both Essentia and MIRToolbox offer a feature describing sensory dissonance (in MIRToolbox, it is called roughness), which is based on the same research of dissonance perception \cite{plomp1965}. We extract this feature and inharmonicity. Inharmonicity only had a weak (0.22) correlation with perceptual dissonance. Figure \ref{fig:results_features} shows a result for the dissonance measure. 
\item \emph{Tonal stability}. HCDF (harmonic change detection function) in MIRToolbox is a feature measuring the flux of a tonal centroid \cite{Harte2006}. This feature was not correlated with our tonal stability feature. 
\item \emph{Modality}. MIRToolbox offers a feature called mode, which is based on an uncertainty in determining the key using pitch-class profiles. 
\end{enumerate}

We could not find features corresponding to melodiousness and rhythmic complexity. Perceptual concepts lack clear definitions, so it is impossible to say that the feature extractor algorithms are supposed to directly measure the same concepts that we had annotated. However, from Figure \ref{fig:results_features} we can see that the chosen descriptors do indeed capture some part of variance in the perceptual features. 

\subsection{Results}

Figure \ref{fig:results_features} shows the results for every mid-feature. For all the mid-features, the best result was achieved by pretraining and fine-tuning the network. Melodiousness, articulation and dissonance could be predicted with a much better accuracy than rhythmic complexity, tonal and rhythmic stability, and mode.

\section{Future Work} 

In this paper, we only investigated seven perceptual features. Other interesting features include tempo, timbre, structural regularity. Rhythmic complexity and tonal stability features had low agreement. It is probable that contributing factors need to be explicitly specified and studied separately. The accuracy could be improved for modality and rhythmic stability. It is not clear whether strong correlations between some features are an artifact of the data selection or music perception. 

\section{Conclusion}

Mid-level perceptual music features could be used for music search and categorization and improve music emotion recognition methods. However, there are multiple challenges in extracting such features: first, such concepts lack clear definitions, and we do not quite understand the underlying perceptual mechanisms yet. In this paper, we collect annotations for seven perceptual features and model them by relying on listener ratings. We provide the listeners with scales with examples instead of definitions and criteria. Listeners achieved good agreement on all the features but two (rhythmic complexity and tonal stability). Using deep learning, we model the features from data. Such an approach has its advantages as compared to specific algorithm-design by being able to pick appropriate patterns from the data and achieve better performance than an algorithm based on a single aspect. However, it is also less interpretable. We release the mid-level feature dataset, which can be used to further improve both algorithmic and data-driven methods of mid-level feature recognition. 

\section{Acknowledgements}

This work is supported by the European Research Council (ERC)  under  the  EUs  Horizon  2020  Framework  Program (ERC Grant Agreement number 670035, project "Con Espressione"). This work was also supported by an FCS grant. 

\end{document}